\documentstyle[amstex,preprint,aps]{revtex}

\begin{document}
\title{Dynamics of Extended Objects: the Einstein-Hilbert Drop}
\author{U. Khanal}
\address{Central Department of Physics, Tribhuvan University, Kirtipur,Kathmandu,
NEPAL}
\date{\today}
\maketitle

\begin{abstract}
The Einstein-Hilbert worldspace action is used to investigate the dynamics
of extended object. In the Robertson-Walker worldspace, this is seen to
introduce a pressureless density which could contribute to dark matter. Such pressureless energy density, present from the very beginning, should have enormous consequences on large scale structure formation in the early Universe.
Generalizing the idea to complexified internal co-ordinates, it becomes possible to gauge the action with U(1) symmetry. A trivial solution of this theory is Einstein's general relativity and source free Maxwell theory. Generally, the equations of motion of the gauge fields are Maxwell equations with source terms that include these fields themselves. The internal co-ordinates, under vacuum domination with negative pressure, obey an EOM that is a hyperbolic wave equation of a charged scalar field that interacts with the gauge fields and gravity in a disperso-conductive medium ; under matter domination with positive pressure however, it is an elliptic potential equation. Since the hyperbolic to
elliptic transition can be made by introducing imaginary time, this result
supports the view that time is actually complex, becoming Minkowskian in
vacuum and Euclidean in matter. A supersymmetric version of the action can also be immediately written down. 

\end{abstract}

\section{Introduction}

In the quest for a unified theory to describe all physical interactions of
nature, the idea of the pointlike fundamental structure of matter is being
abandoned in favour of extended objects now called p-branes. Indeed, most of
the singularities in quantum theory that require renormalization and
effective subtraction of infinities, can be traced to the assumption of
pointlike structure. At present, the one-dimensional string leads the race
towards the theory of everything, leaving the higher dimensional contenders
far behind. But it is just as important to investigate different higher
dimensional incarnations, even if only for elimination, as it is to develop
a viable string theory. A particularly attractive proposition is the
three-dimensional drop evolving with time, as the space-time with which we
are familiar is (3+1) dimensional. Furthermore, both electromagnetic waves
and gravitation cannot propagate in a space of less than three dimensions.
The author has been contending that the Einstein-Hilbert (EH) worldcurvature
action provides a more appropriate and general means, than the Nambu-Goto
(NG) worldvolume action, of studying the gravitodynamics of extended
objects. \cite{uk1} The NG brane that can be considered a special constant
curvature case of the EH one, can also be included as a cosmological term to
constrain the extremization of worldcurvature with simultaneously extremized
worldvolume.

The n+1 dimensional internal co-ordinates $X^{a},0\leq a\leq n,$ are
functions of the p+1 dimensional worldspace co-ordinates $\xi ^{\alpha
},0\leq \alpha \leq p<n.$ $\xi ^{0}=\tau $ is assumed to be timelike, and so
is at least one internal co-ordinate $X^{0}=t.$ The line element $%
ds^{2}=\eta _{ab}dX^{a}dX^{b}=g_{\alpha \beta }d\xi ^{\alpha }d\xi ^{\beta },
$ with the internal space metric tensor $\eta _{ab}$ that need not
necessarily be flat, shows that the induced metric tensor is $g_{\alpha
\beta }=\partial _{\alpha }X^{a}\partial _{\beta }X^{b}\eta _{ab},$where$%
\partial _{\alpha }=\partial /\partial \xi ^{\alpha }.$ The worldspace EH
action is 
\begin{equation}
I=-\frac{1}{2\kappa }\int d^{p+1}\xi \sqrt{g}R,  \label{Eq1}
\end{equation}
where $R$ is the worldcurvature scalar, $g=-\det g_{\alpha \beta }$ and $%
\kappa $ is the gravitational constant appropriate to the p-dimensional
space. When $R$ is constant, Eq.(\ref{Eq1}) is just the NG action. Using the
well known result\cite{weinberg} 
\begin{gather}
\delta I=\frac{1}{2\kappa }\int d^{p+1}\xi \sqrt{g}G^{\alpha \beta }\delta
g_{\alpha \beta }=  \nonumber \\
\frac{1}{2\kappa }\int d^{p+1}\xi \sqrt{g}G^{\alpha \beta }\left[ 2\partial
_{\alpha }X_{a}\partial _{\beta }\delta X^{a}+\partial _{\alpha
}X^{c}\partial _{\beta }X^{b}\frac{\partial \eta _{cb}}{\partial X^{a}}%
\delta X^{a}\right] =0,  \label{Eq1a}
\end{gather}
where $G^{\alpha \beta }$ is the world Einstein tensor. The equation of
motion (EOM), found from the last line of Eq.(\ref{Eq1a}) by performing an
integration by parts of the first term in the square bracket, throwing away
the surface integral with the requirement that the variations $\delta X$
vanish at the boundary, and then demanding that the integrand of the volume
integral be zero for arbitrary $\delta X,$ is 
\begin{equation}
\left( \sqrt{g}\right) ^{-1}\partial _{\alpha }\left[ \sqrt{g}G^{\alpha
\beta }\partial _{\beta }X_{a}\right] -\frac{1}{2}G^{\alpha \beta }\partial
_{\alpha }X^{c}\partial _{\beta }X^{b}\frac{\partial \eta _{cb}}{\partial
X^{a}}=0.  \label{Eq2}
\end{equation}
As $G^{\alpha \beta }$ contains upto second derivatives of the induced
metric, Eq.(\ref{Eq2}) contains upto the fourth derivative of the internal
co-ordinates and is also highly non-linear. Henceforth in this paper, we
will use the locally Minkowskian internal system with $\eta
_{ab}=diag(-1,1,1,1,...),$ whence the second term in the left hand side of
Eq.(\ref{Eq2}) vanishes, making it the same as that discussed in Ref.(1). It
reduces to the wave equation in any worldspace where $G^{\alpha \beta
}\propto g^{\alpha \beta },$ like in the constant curvature worldspace. In
this case, the solutions for $X$ 's are harmonic functions of the $\xi $ 's.
If we further reparametrize the wordspace in co-ordinates $\xi $ that are
themselves harmonic, we may write the EOM as $g^{\alpha \beta }\partial
_{\alpha }\partial _{\beta }X=0.$ In this form, the EOM\ is just of second
order in derivatives, but still non-linear. A trivial solution of Eq.(\ref
{Eq2}), $G^{\alpha \beta }=0,$ are just the Einstein equations for free
space. Depending on the signature of $G^{\alpha \beta },$ particularly the
relative signs of $G^{00}$ and the $G^{ii}$ 's, Eq.(\ref{Eq2}) is generally
found to assume the hyperbolic form if the two signs are different and the
elliptic form if they are the same. These two cases can be interpreted as
vacuum or matter domination respectively. As the hyperbolic to elliptic
transition can be achieved by making time imaginary, this result can be
taken to mean that time is actually complex, becoming Minkowskian in vacuum
and Euclidean in matter. The solution, found in Ref.(1) for a vacuum
dominated case, exhibits an unexpected superluminality due to the dependence
of density on the world space-time. The matter dominated case was also
solved \cite{uk2}. It now appears quite certain that the Universe is endowed
with some form of matter that exerts negative pressure, a prominent
candidate being the cosmological constant that represents the vacuum energy
density of space-time. The extremely high matter pressure of the early
Universe decreases with expansion, and eventually becomes dominated by
however small a cosmological constant there may be. Such a matter-vacuum
transition should have a bearing on some of the unexplained features of the
Universe, particularly in relation to the formation of structures. A further
application of this method to the string with gaussian density and tension 
\cite{uk3} showed that the vibration of the open ends of the string can be
controlled with heavy ends.

In the next Section, this method is applied to the Robertson-Walker(RW)
worldspace. Section III describes a possible method of introducing gauge
interaction in the general relativistic EH brane and looks into some
consequences. The final Section makes some concluding remarks on the results
and points towards some directions in which the theory can be further
developed.

\section{Robertson-Walker drop}

\bigskip In the first step towards the interacting EH drop, we generalize
the action of Eq.(\ref{Eq1}) to 
\begin{equation}
J=-\frac{1}{2\kappa }\int d^{p+1}\xi \sqrt{g}R-\rho _{V}\int d^{p+1}\xi 
\sqrt{g}+I_{M}  \label{Eq3}
\end{equation}
\noindent where the cosmological term involving the vacuum energy density $%
\rho _{V}$ is just the NG part and $I_{M}$ is the action of matter.
Requiring $J$ to be stationary with respect to variations of the internal
co-ordinates gives $\delta J=\frac{1}{2\kappa }\int d^{p+1}\xi \sqrt{g}\left[
G^{\alpha \beta }-\kappa \left( \rho _{V}g^{\alpha \beta }-T_{M}^{\alpha
\beta }\right) \right] \delta g_{\alpha \beta }=\frac{1}{\kappa }\int
d^{p+1}\xi \sqrt{g}\left[ G^{\alpha \beta }-\kappa \left( \rho _{V}g^{\alpha
\beta }-T_{M}^{\alpha \beta }\right) \right] \partial _{\alpha }\delta
X^{a}\partial _{\beta }X_{a}=0.$ Integrating the last expression by parts
and then discarding the surface integral leads us to the EOM, 
\begin{equation}
\left( \sqrt{g}\right) ^{-1}\partial _{\alpha }\left[ \sqrt{g}h^{\alpha
\beta }\partial _{\beta }X_{a}\right] =0,  \label{Eq4}
\end{equation}
where $h^{\alpha \beta }=G^{\alpha \beta }-\kappa \rho _{V}g^{\alpha \beta
}-T_{M}^{\alpha \beta },$ and $T_{M}^{\alpha \beta }$ is the energy-momentum
tensor of matter .  Generally, Eq.(\ref{Eq4}) is fourth order non-linear
partial differential equation that will prove to be quite difficult to
solve. But a trivial solution, $h^{\alpha \beta }=0,$ just reproduces the
Einstein field equations $G^{\alpha \beta }-\kappa \rho _{V}g^{\alpha \beta
}=-\kappa T_{M}^{\alpha \beta }.$ Thus we can interpret $h_{\alpha \beta }$
as a tensor that describes the deviation of the spacetime from that of
general relativity.

To look into some simple consequences, let us identify the internal and
world times by requiring $\tau =\xi ^{0}=X^{0}=t$, in which case Eq.(\ref
{Eq4}) for $X^{0}$ becomes $\partial _{\alpha }\left[ \sqrt{g}h^{\alpha 0}%
\right] =0.$ If $h^{\alpha \beta }$ is a diagonal tensor, then the solution
is $h^{00}=f(r,\theta ,\phi )/\sqrt{g}$ . In RW worldspace which is
homogeneous and isotropic, the solution is $R^{3}h_{\tau }^{\tau }=f({\bf %
\xi )}$ where $R$ is the scale factor. As the left hand side is only $\tau $
dependent, $f({\bf \xi })$ has to be a constant which can be written as $%
\frac{\kappa }{3}\rho _{B}\left( R_{0}\right) ^{3}.$ Thus we can write the
energy conservation equation in the usual Friedmann form as 
\begin{equation}
\left( \frac{\dot{R}}{R}\right) ^{2}+\frac{k}{R^{2}}=\frac{\kappa }{3}\left[
\rho _{V}+\rho _{M}+\rho _{B}\left( \frac{R_{0}}{R}\right) ^{3}\right] 
\label{Eq4a}
\end{equation}
with $\dot{R}=\partial R/\partial \tau ,$ whence $\rho _{B}$ can be
understood as a dusty, pressureless, uniform background density contributed
by the internal co-ordinates.The space-space part $h_{j}^{i}=\left[ 2\frac{%
\ddot{R}}{R}+\left( \frac{\dot{R}}{R}\right) ^{2}+\frac{k}{R^{2}}+\frac{%
\kappa }{3}\left( P_{M}+P_{V}\right) \right] g_{j}^{i},$ \noindent \noindent
where $P_{M}$ is the matter pressure and $P_{V}=-\rho _{V}$ is the vacuum
pressure, can be shown to vanish by differentiating $R^{3}\times Eq.(\ref
{Eq4a})$ with respect to $\tau $ to determine $\ddot{R},$ using the
energy-momentum conservation equation $\partial \left( R^{3}\rho \right)
/\partial R=-3PR^{2},$ and making appropriate substitution from Eq.(\ref
{Eq4a}). Then the other independent equation also takes on the familiar form 
\begin{equation}
\frac{\ddot{R}}{R}=-\frac{\kappa }{6}\left[ 3P+\rho \right] ;  \label{Eq4b}
\end{equation}
here the total density is $\rho =\rho _{M}+\rho _{V}+\rho _{B}R_{0}^{3}/R^{3}
$ and the total pressure $P=P_{M}-\rho _{V}$ as the term with $\rho _{B}$
does not contribute any pressure. 

Such a background $\rho _{B}$ could constitute a cold dark matter that was
present from the very beginning. The presence of this pressureless energy
density that behaves as cold dark matter from the initial time, should play
a very important role in the evolution of the Universe, and on structure
formation. Without the presence of pressure to counteract gravity, this
energy density could easily undergo gravitational collapse from the outset.
Such regions in space could form the seeds around which other matter accrete
to form the structures we see. 

Under these conditions the EOM of the extra internal co-ordinates, Eq.(\ref
{Eq4}), are found to be $\partial _{\tau }^{2}X_{r}=0,r>3.$ These have
solutions that are linear in $\tau ,$ viz., 
\begin{equation}
X_{r}(\tau ,{\bf \xi })=A_{r}({\bf \xi })+B_{r}({\bf \xi })\times \tau ,
\label{Eq4c}
\end{equation}
where $A$ and $B$ are arbitrary functions of spatial co-ordinates that have
to be fixed by boundary conditions. 

\section{Interacting Einstein-Hilbert drop}

To introduce gauge field interaction into the theory, we have to first
complexify the internal co-ordinates $X$ and write $g_{\alpha \beta }=\frac{1%
}{2}\left[ \partial _{\alpha }X^{\ast a}\partial _{\beta }X_{a}+\left(
\alpha \leftrightarrow \beta \right) \right] ,$ where $X^{\ast a}$ are
complex conjugates of $X^{a}.$ We are motivated to complexify the internal
co-ordinates by the discussion in Sec.I that the time (and also space)
co-ordinate may actually be complex. Under a global phase change of $X$ by a
factor $e^{iq\phi },$ where $q$ is the charge and $\phi $ is a constant
phase angle, $g_{\alpha \beta }$ remains invariant and so does $J,$ provided 
$I_{M}$ is also invariant. This invariance will generate a conserved charge
current.

To make the theory invariant under a local U(1) transformation when $\phi $
is a function of the $\xi ^{\alpha }$ 's, we can use our knowledge of
electrodynamics to introduce real gauge potentials $A_{\alpha }$ with
minimal coupling to write 
\begin{equation}
g_{\alpha \beta }=\frac{1}{2}\left[ \left( \partial _{\alpha }X^{\ast
a}+iqA_{\alpha }X^{\ast a}\right) \left( \partial _{\beta }X_{a}-iqA_{\beta
}X_{a}\right) +\left( \alpha \leftrightarrow \beta \right) \right] .
\label{Eq5}
\end{equation}
It can be easily checked that $g_{\alpha \beta }$ is invariant under the
simultaneous gauge transformation 
\begin{align*}
X& \rightarrow X^{\prime }=e^{iq\phi }X, \\
A_{\alpha }& \rightarrow A_{\alpha }^{\prime }=A_{\alpha }+\partial _{\alpha
}\phi .
\end{align*}
Explicitly, $\partial _{\alpha }X^{\prime }-iqA_{\alpha }^{\prime }X^{\prime
}=e^{iq\phi }\left[ \partial _{\alpha }X-iqA_{\alpha }X\right] ,$ and the
phase factor will vanish on multiplying by the conjugate. Hence, any
quantity calculated from $g_{\alpha \beta },$ like $g$ and $R,$ are
invariant. The field strength tensor $F_{\alpha \beta }=\partial _{\alpha
}A_{\beta }-\partial _{\beta }A_{\alpha }$ is also seen to be invariant
under the gauge transformation. For a gauge invariant action, we have to
include the Lagrangian due to the gauge fields, $-\frac{1}{4}F_{\alpha \beta
}F^{\alpha \beta },$ and modify the action $J$ of Eq.(\ref{Eq3})to 
\begin{equation}
K=-\frac{1}{2\kappa }\int d^{p+1}\xi \sqrt{g}R-\rho _{V}\int d^{p+1}\xi 
\sqrt{g}-\frac{1}{4}\int d^{p+1}\xi \sqrt{g}F_{\alpha \beta }F^{\alpha \beta
}+I_{M}.  \label{Eq6}
\end{equation}
Imposing the stationarity of this action, we have 
\begin{equation}
\delta K=\frac{1}{2\kappa }\int d^{p+1}\xi \sqrt{g}H^{\alpha \beta }\delta
g_{\alpha \beta }-\int d^{p+1}\xi \sqrt{g}F^{\alpha \beta }\partial _{\alpha
}\delta A_{\beta }=0,  \label{Eq7}
\end{equation}
where $H^{\alpha \beta }=G^{\alpha \beta }-\kappa \left( \rho _{V}g^{\alpha
\beta }-T_{M}^{\alpha \beta }-T_{F}^{\alpha \beta }\right) $ has been
modified from $h^{\alpha \beta }$ of Eq.(\ref{Eq4}) to include the
energy-momentum tensor of the gauge field $T_{F}^{\alpha \beta }=F^{\alpha
}\,_{\mu }F^{\beta \mu }-\frac{1}{4}g^{\alpha \beta }F_{\mu \nu }F^{\mu \nu
}.$ Now, $\delta g_{\alpha \beta }=\frac{1}{2}[\left( \partial _{\alpha
}\delta X^{\ast a}+iqA_{\alpha }\delta X^{\ast a}+iq\delta A_{\alpha
}X^{\ast a}\right) \left( \partial _{\beta }X_{a}-iqA_{\beta }X_{a}\right)
+\left( \partial _{\alpha }X^{\ast a}+iqA_{\alpha }X^{\ast a}\right) \left(
\partial _{\beta }\delta X_{a}-iqA_{\beta }\delta X_{a}-iq\delta A_{\beta
}X_{a}\right) +\left( \alpha \leftrightarrow \beta \right) ].$ For the EOM
of $X$ we take the variations in Eq.(\ref{Eq7}) to be $\delta X^{\ast }\neq
0,$ $\delta X=\delta A_{\alpha }=0,$ perform an integration by parts and
discard the surface integral to find 
\begin{equation}
\left( \sqrt{g}\right) ^{-1}\left( \partial _{\alpha }-iqA_{\alpha }\right) %
\left[ \sqrt{g}H^{\alpha \beta }\left( \partial _{\beta }-iqA_{\beta
}\right) X_{a}\right] =0.  \label{Eq8}
\end{equation}
Obviously, \noindent $X^{\ast }$ will satisfy the conjugate of Eq.(\ref{Eq8}%
). Next, for the EOM of $A_{\alpha },$ we take the variations $\delta
X^{\ast }=$ $\delta X=0,\delta A_{\alpha }\neq 0.$ Then an integration by
parts of the last integral in Eq.( \ref{Eq7}) and throwing away the surface
part leaves us with the inhomogeneous Maxwell equations 
\begin{equation}
\left( \sqrt{g}\right) ^{-1}\partial _{\alpha }\left[ \sqrt{g}F^{\alpha
\beta }\right] =-\frac{1}{\kappa }H^{\alpha \beta }j_{\alpha }  \label{Eq9}
\end{equation}
where the p+1 charge current of the internal co-ordinates are $j_{\alpha }=-%
\frac{iq}{2}\left[ \left( \partial _{\alpha }X^{\ast a}+iqA_{\alpha }X^{\ast
a}\right) X_{a}-X^{\ast a}\left( \partial _{\alpha }X_{a}-iqA_{\alpha
}X_{a}\right) \right] .$ The homogeneous Maxwell equations, 
\begin{equation}
\partial _{\alpha }F_{\beta \gamma }+\partial _{\beta }F_{\gamma \alpha
}+\partial _{\gamma }F_{\alpha \beta }=0,  \label{Eq10}
\end{equation}
are automatically satisfied from the definition of $F_{\alpha \beta }.$ So
the action $K$ generates both gravitation and electromagnetism.

As in the non-interacting case of Eq.(\ref{Eq4}), a trivial solution of Eq.(%
\ref{Eq8}), $H^{\alpha \beta }=0,$ is just the Einstein field equations 
\begin{equation}
G^{\alpha \beta }-\kappa \rho _{V}g^{\alpha \beta }=-\kappa \left(
T_{M}^{\alpha \beta }+T_{F}^{\alpha \beta }\right) .  \label{Eq11}
\end{equation}
In this case, Eq.(\ref{Eq9}) becomes source-free. Thus, general relativity
is included in this theory of interacting EH drop. General solution of Eq.(%
\ref{Eq8}) should describe new effects. Another trivial solution of Eq.(\ref
{Eq8}) can be used to write the gauge fields as one of the internal
co-ordinates. In particular, $\left( \partial _{\beta }-iqA_{\beta }\right)
X_{n}=0$ is solved by 
\begin{equation}
X_{n}=L_{n}\exp \left[ iq\int A_{\alpha }d\xi ^{\alpha }\right] ,
\label{Eq12}
\end{equation}
where $L_{n}$ is a constant. In essence, this is an equivalent way of
looking at the Kaluz-Klein idea of getting electromagnetism from extra
dimension. Another simple solution of Eq.(\ref{Eq8}), 
\begin{equation}
H^{\alpha \beta }\left( \partial _{\beta }-iqA_{\beta }\right) X_{a}=0,
\label{Eq13}
\end{equation}
is solved by all $X_{a}$ 's having the form of Eq.(\ref{Eq12}) for all
non-singular $H^{\alpha \beta }.$

For other solutions of Eq.(\ref{Eq8}) we particularly need a knowledge of $%
T_{M}^{\alpha \beta }.$ As an example, let us consider a worldspace of
constant curvature with $G^{\alpha \beta }=-\frac{p(p+1)}{2}Cg^{\alpha \beta
}=-\kappa \rho _{C}g^{\alpha \beta }$ where $C$ is the curvature constant
and $\rho _{C}=\frac{p(p+1)}{2\kappa }C$ may be considered the energy
density due to curvature. Isotropic, perfect fluid matter in its co-moving
co-ordinates has $T_{M\quad \beta }^{\quad \alpha }=diag(-\rho
_{M},P_{M},P_{M},...).$ Assuming the electric and magnetic fields to be
isotropic as well, we may write \noindent $T_{F\quad \beta }^{\quad \alpha
}=diag(-\rho _{F},\rho _{F}/p,\rho _{F}/p,\rho _{F}/p,...)$ where the
radiation pressure is $P_{F}^{\quad }=\rho _{F}/p.$ Specializing further to
the $p=3$ case, we find that $H_{\quad \beta }^{\alpha }$\noindent $=\kappa
\quad diag(-\rho ,P,P,P),$ where the total density is $\rho =\rho _{V}+\rho
_{C}+\rho _{M}+\rho _{F}$ and the total pressure is $P=P_{M}+\rho
_{F}/3-\rho _{V}-\rho _{C}.$ Then Eq.(\ref{Eq9}) can be explicitly written
down in terms of the familiar electric and magnetic field vectors ${\bf E}$
and ${\bf B}$ and the charge four current of the internal co-ordinates $%
j^{\alpha }=\left( j^{0},{\bf j}\right) $ as 
\begin{align}
& \nabla .{\bf E=-}\rho j^{0}\text{ and}  \label{Eq14} \\
& \nabla {\bf \times B-}\frac{1}{\sqrt{g}}\partial _{\tau }\left( \sqrt{g}%
{\bf E}\right) ={\bf \allowbreak }P{\bf j}\newline
.  \label{Eq15}
\end{align}
Eq.(\ref{Eq14}) shows coupling of the charge density of the internal
co-ordinates $j^{0}$ with the total energy density at that point to produce
the source for the electric field. Hence, the electric field contributes to
its own source through its energy density $\rho _{F}=\left(
E^{2}+B^{2}\right) /2,$ making the Poisson equation, Eq.(\ref{Eq14}),
non-linear. Similar is the case with Ampere's law, Eq.(\ref{Eq15}), where
the magnetic field also contributes to its own source. In this simplified
scenario, Eq.(\ref{Eq8}) can be written as 
\begin{equation}
\frac{1}{\sqrt{g}}\left( \frac{\partial }{\partial \tau }-iqA_{0}\right) %
\left[ \sqrt{g}{\small \rho }\left( \frac{\partial }{\partial \tau }%
-iqA_{0}\right) X\right] +\left( \nabla -iq{\bf A}\right) .\left[ P\left(
\nabla -iq{\bf A}\right) X\right] =0.  \label{Eq16}
\end{equation}
In vacuum with $\rho =-P=$ positive constant, Eqs.(\ref{Eq14})-(\ref{Eq16})
are exactly the same as those for the electrodynamics of a massless scalar
field. Assuming positive energy density, Eq.(\ref{Eq16}) is hyperbolic under
vacuum domination with negative pressure and elliptic under matter
pomination with positive pressure. Thus in the general case where $\rho $
and $P$ are both spacetime dependent, the theory described here for the
bosonic internal co-ordinates, is found to generalize scalar electrodynamics
to a disperso-conductive medium.

A supersymmetric version of the theory, capable of generating fermionic
interaction as well, can be written down just by supersymmetrizing the
worldspace metric to $g_{\alpha \beta }=\frac{1}{2}\left[ \overline{\Pi }%
_{\alpha }{}^{a}\Pi _{\beta }{}^{b}\eta _{ab}+\left( \alpha \leftrightarrow
\beta \right) \right] $ with $\Pi _{\alpha }{}^{a}=\left( \partial _{\alpha
}-iqA_{\alpha }\right) X^{a}+\overline{\theta }\Gamma ^{a}\left( \partial
_{\alpha }-iqA_{\alpha }\right) \theta $ where $\theta $ are the fermionic
co-ordinates and the $\Gamma $ 's are the n+1 dimensional Dirac matrices.
The possibility of gauging such a supersymmetric theory with U(1), or with even larger symmetries like SU(n+1) or a n+1dimensional Poincar\`{e} type is also open. The mechanism for an effective breaking of this large symmetry into a 3+1 Lorentz group, low energy representation of the remaining components, the verifiable predictions, dimensionality of the internal space, the results of
quantization and supersymmetrization, etc., are some open problems.

\section{Conclusions}

In this paper, some results of generalizing the brane action from NG to EH
action of the worldspace has been investigated. The NG action is contained
as a special, constant-curvature, case in this general relativistic theory.
The NG part can also be included in the general action as an effective
cosmological term, with the cosmological constant appearing as a Lagrange
multiplier that constrains the extremization of the curvature simultaneously
with extremized volume. Applying this method to the RW space gives an EOM
that is exactly similar to what we are familiar with, except that a
pressureless matter is introduced. This pressureless background density,
contributed by the internal co-ordinates that grow linearly with time, is
present from the very beginning. So it could account significantly as a cold
dark matter that is present throughout the evolution of the Universe.

Complexifying the internal co-ordinates allows us to introduce the charge
and consider interaction amongst the internal co-ordinates. The knowledge of
electrodynamics was used as hindsight to gauge this theory of interacting EH
drop with U(1) symmetry. The equations of motion\ that result are
generalized partial differential equation of either the hyperbolic or
elliptic type for the internal co-ordinates, and Maxwell equations with the
gauge field self-interaction contributions to the source terms for the gauge
fields. Under simplifying assumptions like constant-curvature worldspace and
isotropic energy-momentum tensors of matter and electromagnetic field in $p=3
$ worldspace, the EOM of the internal co-ordinates reduce to very familiar
form that combines two important equations of physics into one. In a
worldspace that is dominated by vacuum, with the total pressure negative,
the equations are just those of scalar electrodynamics, albeit, modified a
little with temporal damping due to time dependence of the density, and
spatial dispersion due to pressure gradient. The internal co-ordinates
behave as charged, massless scalar field in a disperso-conductive medium,
interacting with the electromagnetic field and gravity. Under matter
domination with positive total pressure, this EOM becomes an elliptic type.
As the hyperbolic to elliptic transition can be made by making the worldtime
imaginary, these results further support the author's proposition that time
is in fact complex, becoming Minkowskian under vacuum domination and
Euclidean under matter domination. In an earlier investigation of the
non-interacting EH drop in Ref.(1), the author had shown that the
conductivity that arises due to the time dependence of the density of the
medium interacts with the internal co-ordinate field to give it a tachyonic
behaviour. When the worldtime is made imaginary, the conductivity also
becomes imaginary. Thus the imaginary conductivity will behave as mass of
the scalar field. These conclusions also hold in the theory of the
interacting general relativistic drop presented in this paper. In the
complexified worldtime plane, the real axis represents the vacuum dominated
Minkowskian time, while the imaginary axis represents matter dominated
Euclidean time.

Another important result is that the EOM of this theory, Eq.(\ref{Eq8}), has
a trivial solution Eq.(\ref{Eq11}) which is general relativity. In this case
the Maxwell equations become source-free, indicating that general relativity
is the vacuum solution of this theory of interacting EH drop. Application of
this theory to any other situation besides this trivial case will describe
physics at a level that is wider than general relativity and
electrodynamics. Another trivial solution can be used to relate one of the
internal co-ordinates to the gauge potentials. Supersymmetric generalization
is also immediate, but the necessary Wess-Zumino and other terms, as well as
the consequences have to be worked out. Only the classical theory has been
described here, so its consistency under quantization, as well as the
quantum effects, have to be investigated.

In this paper, the theory has just been gauged with U(1) symmetry. It is
also possible to gauge it with any other relevant group like SU(n+1) or a
generalized n+1 dimensional Poincar\`{e} type group. The symmetry breaking
mechanism, resultant low energy representations, and further predictions of
this theory will be investigated in future work.

\end{document}